# Expected spectral characteristics of (101955) Bennu and (162173) Ryugu, targets of the OSIRIS-REx and Hayabusa2 missions

J. de León[(1,2)][1], H. Campins[(3)], D. Morate[(1,2)], M. De Prá[(4)], V. Alí-Lagoa[(5)], J. Licandro[(1,2)], J. L. Rizos[(1,2)], N. Pinilla-Alonso[(6)], D. N. DellaGiustina[(7)], D. S. Lauretta[(7)], M. Popescu[(1,2)], V. Lorenzi[(9,1)]

(1) Instituto de Astrofísica de Canarias, C/Vía Láctea s/n, E-38205 La Laguna, Tenerife, Spain
(2) Departamento de Astrofísica, Universidad de La Laguna, E-38206 La Laguna, Tenerife, Spain
(3) Physics Department, University of Central Florida, P.P. box 162385, Orlando, FL 32816-2385, USA
(4) Departamento de Astrofísica, Observatorio Nacional, Rio de Janeiro, 20921-400, Brazil
(5) Max-Planck-Institut für extraterrestrische Physik, Giessenbachstrasse 1, 85748 Garching, Germany
(6) Florida Space Institute, University of Central Florida, Orlando, FL 32816, USA
(7) Lunar and Planetary Laboratory, University of Arizona, Tucson, AZ, USA
(8) Fundación Galileo Galilei – INAF, Rambla José Ana Fernández Pérez, 7, E-38712 Breña Baja, La Palma, Spain

**Abstract**

NASA's OSIRIS-REx and JAXA's Hayabusa2 sample-return missions are currently on their way to encounter primitive near-Earth asteroids (101955) Bennu and (162173) Ryugu, respectively. Spectral and dynamical evidence indicates that these near-Earth asteroids originated in the inner part of the main belt. There are several primitive collisional families in this region, and both these asteroids are most likely to have originated in the Polana-Eulalia family complex. We present the expected spectral characteristics of both targets based on our studies of four primitive collisional families in the inner belt: Polana-Eulalia, Erigone, Sulamitis, and Clarissa. Observations were obtained in the framework of our PRIMitive Asteroids Spectroscopic Survey (PRIMASS). Our results are especially relevant to the planning and interpretation of in situ images and spectra to be obtained by the two spacecraft during the encounters with their targets.

## 1. Introduction

Near-Earth asteroids (NEAs) are among the most interesting populations of minor bodies of the Solar System: their proximity to the Earth makes them impact hazards and also accessible to spacecraft. Consequently, they can be studied in detail to mitigate potential impacts and also sample their surface material for analysis in terrestrial laboratories. Also, NEAs are considered ideal targets for In-Situ Resource

---

[1] e-mail: jmlc@iac.es

Utilization (ISRU) and could become a source of materials for space activities in the near future. Among NEAs, those with a primitive composition are of particular interest, as they might contain water and organic compounds, being the remnants of the formative stages of our Solar System and thereby providing information on the early conditions of the solar nebula.

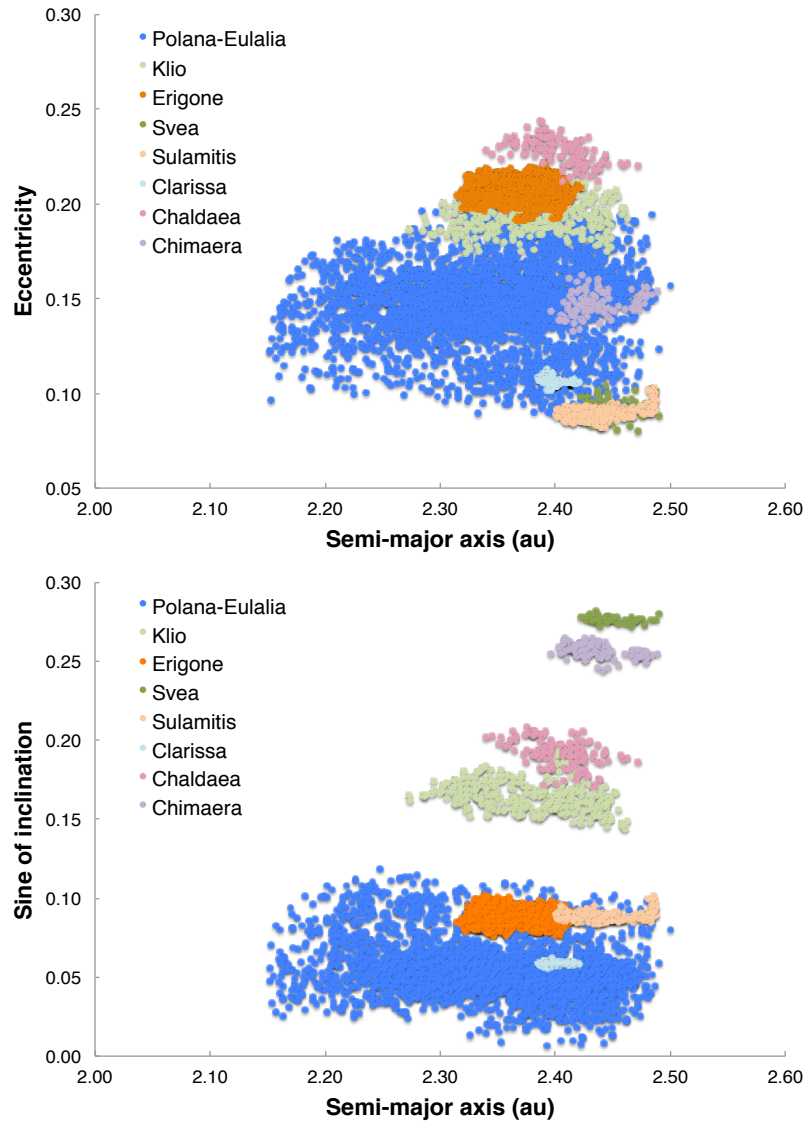

**Figure 1**. Proper semi-major axis vs. proper eccentricity (upper panel) and sine of proper inclination (bottom panel) for the currently identified eight primitive collisional families located in the inner belt, according to Nesvorný et al. (2015). For the Polana-Eulalia family complex we have used the definition from Walsh et al. (2013).

NEA lifetimes are short compared to the age of the Solar System (Morbidelli et al. 2002), and there must exist a replenishment mechanism to sustain the population. Dynamical models indicate that most of the NEAs come from the main asteroid belt (Bottke et al. 2002), in particular from the region enclosed by the $\nu_6$ secular resonance with Jupiter and Saturn, located at 2.1 au, and the 3:1 mean motion resonance with Jupiter, at 2.5 au. We refer to this region as the inner main belt. The process that best describes how NEAs reach near-Earth space can be separated into two steps. First, collisions in the main belt generate small fragments in the size range of NEAs, i.e., meters to a few kilometers. Second, the action of the Yarkovsky effect, which is more efficient for small-diameter objects, modifies the semi-major axes of the orbits of these fragments until they reach one of the transport routes above mentioned (Morbidelli & Vokrouhlický 2003; Bottke et al. 2006). Therefore, collisional families in the inner main belt are considered the most likely source of NEAs. This rationale also applies to primitive NEAs that are targets of space missions. As of early 2018, there are eight primitive collisional families identified in this region (Nesvorný et al. 2015). These are, the Polana-Eulalia complex, and the Erigone, Sulamitis, Clarissa, Klio, Chaldaea, Svea, and Chimaera collisional families (Fig. 1).

There are currently two sample-return missions on their way to two primitive NEAs: NASA's OSIRIS-REx mission to asteroid (101955) Bennu (Lauretta et al. 2017) and JAXA's Hayabusa2 mission to asteroid (162173) Ryugu (Tsuda et al. 2013). Asteroid Bennu is classified as a B-type asteroid from its visible spectrum (Clark et al. 2011), and its most likely origin is the Polana-Eulalia family complex (Campins et al. 2010; Walsh et al. 2013; Bottke et al. 2015). Likewise, Ryugu, a C-type asteroid, most likely originated in the Polana-Eulalia complex or in the population of low-albedo and low-inclination background asteroids (Campins et al. 2013), identified for the first time by Gayon-Markt et al. (2012). With the aim of enhancing the science return of both missions, we started our PRIMitive Asteroids Spectroscopic Survey (PRIMASS) in 2010. As part of the PRIMASS project, we obtain visible and near-infrared spectra of the members of the collisional families and dynamical groups of the main asteroid belt, as well as other populations of primitive objects (De Prá et al. 2018). As of 2018, we have characterized four primitive families in the inner belt: Polana-Eulalia, Erigone, Sulamitis, and Clarissa. We are in the process of data reduction and analysis of the visible spectra of members of the rest of the primitive families in the inner belt (Klio, Chaldaea, Svea, and Chimaera).

Our results on the spectral characterization of the members of the Polana-Eulalia family complex both in the visible (de León et al. 2016) and the near-infrared (Pinilla-Alonso et al. 2016) provide essential context for this study. Despite the dynamical complexity of this region (Walsh et al. 2013; Milani et al. 2014; Dykhuis & Greenberg 2015), asteroids belonging to this complex show spectral homogeneity, with a continuum of spectral slopes from blue to moderately red, typical of B- and C-type objects. In contrast, Morate et al. (2016) showed that the Erigone collisional family shows a more extensive spectral diversity, with B-, C-, X-, and T-type objects. Also, the majority of the Erigone family members have an absorption band at 0.7 µm, as does the parent body (163) Erigone; this absorption is associated with aqueously altered silicates, i.e., phyllosilicates (Vilas 1994). Neither (142) Polana nor (495) Eulalia or their family members show this band. Finally, our most recent results on the Sulamitis and the Clarissa families suggest that there might be a connection between Polana-Eulalia and Clarissa (both families with no hydration signatures and presenting mainly B- and C-type asteroids). Our work revealed a similar association between Erigone and Sulamitis (both families showing the 0.7 µm feature and a similar fraction of B-, C-, X-, and T-type asteroids). This tentative connection has been presented in Morate et al. (2018) and will be further explored in this work.

Visible spectra of the parent bodies of the collisional families, as well as their members are presented in Section 2, together with a summary of the results obtained for the spectral analysis of each family. In Section 3 we compare the available visible spectra of both Bennu and Ryugu with the spectra of the B- and C-type asteroids found in the Polana-Eulalia complex and also in the four collisional families of primitive asteroids of the inner belt studied in this work. In the case of Bennu, we also compute (*b'-v*) and (*v-x*) colors using the response curves of the MapCam filters (one of the three scientific cameras on-board OSIRIS-REx: Rizk et al. 2018) and the mean spectra of the different taxonomies mentioned before. Section 4 presents a discussion of the obtained results, and conclusions are presented in Section 5.

## 2. Collisional families of primitive asteroids in the inner belt

The primary aim of this work is to make a comparative study between the ground-based visible spectra of the primitive NEAs (101955) Bennu and (163172) Ryugu, and the visible spectra of the primitive collisional families in the inner belt, which are considered as the most likely source region of these NEAs. By studying the spectral properties of the members of these families, we can constrain the expected global spectral properties of the surface of the targets before the arrival of the spacecraft. In this section, we study the spectral characteristics of the parent bodies and the family members.

### 2.1. Parent bodies

In Fig. 2 we present the available visible spectra of the parent bodies of the primitive collisional families studied in this work: asteroids (142) Polana, (495) Eulalia, (163) Erigone, (752) Sulamitis, and (306) Clarissa. For those asteroids having more than one

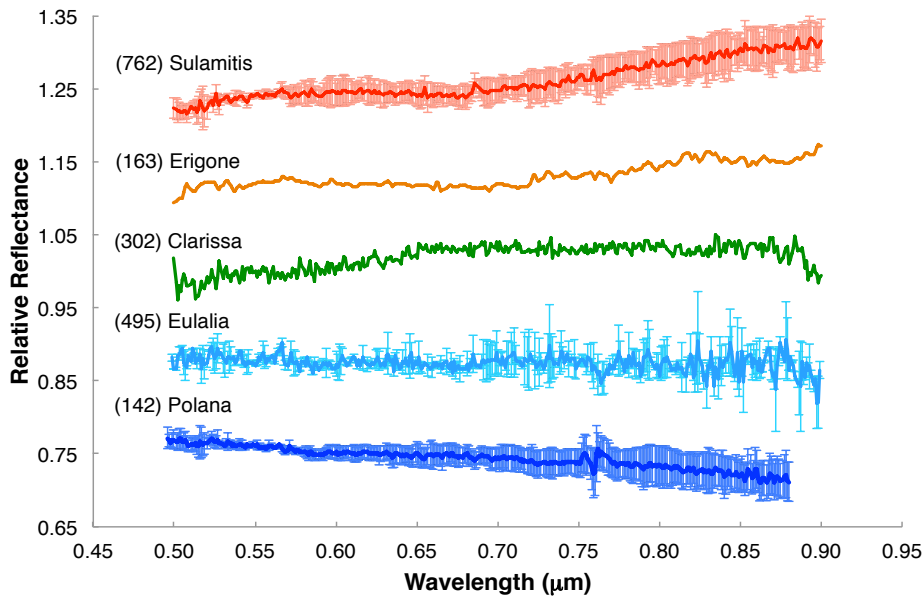

**Figure 2**. Visible spectra of the parent bodies of the primitive collisional families studied in this work. The spectra are normalized to unity at 0.55 μm and offset vertically for clarity. Asteroids have been ordered according to their spectral slope, from the reddest object at the top (Sulamitis) to the bluest asteroid at the bottom (Polana). See the main text for details on individual spectra.

 published visible spectrum we compute the average spectrum and the corresponding ±1σ deviation of the mean (shown as error bars). Table 1 provides a summary of the observational circumstances for each parent body, including the observation date, the phase angle (α) at the time of observation, the telescope used, and the corresponding bibliographic reference. In a separate table (Table 2) we show some physical properties of these asteroids, including size and visible geometric albedo ($p_V$). The table also includes the computed values for the spectral slope $S'$. This slope has been computed in the range from 0.55 to 0.88 μm (this is the common wavelength interval to the spectra of the five asteroids), using the expression as defined by Luu & Jewitt (1990), $S' = (dS/d\lambda)/S_{0.55}$, where $dS/d\lambda$ is the variation of the reflectance in the selected wavelength range, and $S_{0.55}$ is the reflectance at 0.55 µm. For those asteroids having more than one published spectrum we see no correlation between phase angle and spectral slope (these are 142 Polana, 495 Eulalia, and 752 Sulamitis).

**Table 1**. Summary of the observational circumstances of the parent bodies of the primitive families studied in this work.

| Parent body | Obs. Date | Phase angle (°) | Telescope | Ref. |
|---|---|---|---|---|
| (142) Polana | 25/04/1996 | 8.1 | 1.3m McG-H | B02 |
| | 19/04/1992 | 14.5 | 1.5m CTIO | V92 |
| | 16/04/1999 | 23.1 | 1.52m ESO | F14 |
| | 20/07/2011 | 27.9 | 3.6m NTT | dL16 |
| (495) Eulalia | 06/07/1987 | 18.3 | 1.5m CTIO | V92 |
| | 10/09/1992 | 9.8 | 2.4m H | X95 |
| | 05/05/2014 | 13.7 | 3.6m TNG | dL16 |
| (302) Clarissa | 31/08/2016 | 22.6 | 10.4m GTC | M17 |
| (163) Erigone | 21/04/1996 | 10.4 | 1.3m McG-H | B02 |
| (752) Sulamitis | 29/11/1992 | 9.1 | 2.4m H | X95 |
| | 26/01/2001 | 17.4 | 1.52m ESO | L04 |
| | 21/07/2015 | 23.9 | 2.5m INT | M17 |
| | 31/08/2015 | 19.6 | 10.4m GTC | M17 |

*Telescopes*: McG-H – McGraw-Hill (Kitt Peak, Arizona); CTIO – Cerro Tololo Inter-american Observatory (La Serena, Chile); ESO – European Southern Observatory (La Silla, Chile); H – Hiltner (Kitt Peak, Arizona); INT – Isaac Newton Telescope (El Roque de Los Muchachos, Spain); TNG – Telescopio Nazionale Galileo (El Roque de los Muchachos, Spain); NTT – New Technology Telescope (La Silla, Chile); GTC – Gran Telescopio Canarias (El Roque de Los Muchachos, Spain).
*Ref*: B02 – Bus & Binzel (2002); dL16 – de León et al. (2016); F14 – Fornasier et al. (2014); L04 – Lazzaro et al. (2004); M17 – Morate et al. (2018); V92 – Vilas & McFadden (1992); X95 – Xu et al. (1995)

The slope is computed by a simple linear fit in the 0.55-0.88 µm range, in units of %/1000 Å. Typically, the most significant contribution to the error in the spectral slope comes from the process of dividing the spectrum of the asteroid by the spectra of solar analog stars. This error is usually lower that 1.0 %/1000 Å and is typically of the order of 0.5 %/1000 Å. As we do not have access to this information for all the targets, we show in Table 2 the error associated with the slope computation process: we perform 100 iterations, randomly removing 10% of the data points and doing a linear fitting on each iteration. The resulting slope is the mean of these 100 iterations, and the error is the standard deviation of this mean.

**Table 2**. Physical properties of the parent bodies of the primitive families studied in this work.

| Parent body | Size[1] (km) | $p_V$[1] | $S'$ (%/1000Å) |
|---|---|---|---|
| (142) Polana | 58 ± 6 | 0.044 | -1.17 ± 0.01 |
| (495) Eulalia | 38 ± 4 | 0.054 | -0.11 ± 0.02 |
| (302) Clarissa | 39 ± 4 | 0.046 | 1.00 ± 0.02 |
| (163) Erigone | 76 ± 7 | 0.041 | 1.25 ± 0.02 |
| (752) Sulamitis | 61 ± 6 | 0.040 | 2.17 ± 0.02 |

[1] Average diameter computed using the NEATM model from Alí-Lagoa et al. (2018) to fit NEOWISE and AKARI data. The corresponding visible geometric albedos used the $H$-$G_{12}$ values from Oszkiewicz et al. (2012).

We see from the spectra in Fig. 2 and the slopes in Table 2 that there is a significant variation in the spectral properties of these parent bodies, even if they are all considered primitive asteroids. Their visible spectra range from featureless and blue sloped (Polana) to neutral colors (Clarissa) and red sloped (Sulamitis and Erigone) with a clear absorption band at 0.7 µm associated with phyllosilicates and indicative of aqueous alteration processes. Interestingly, we also observe a wide range in the estimated ages of their corresponding collisional families (Table 3). Also, the observed spectral diversity among the parent bodies is even more extensive in that of the members of these families; we discuss this point in detail in the next section.

## 2.2. Family members

We have spectrally characterized four out of the eight primitive collisional families in the inner belt: Polana-Eulalia, Erigone, Sulamitis, and Clarissa. We have computed the spectral slopes not only for the parent bodies (Table 2) but also for all the members observed within the families. The results can be found in de León et al. (2016) and Morate et al. (2016, 2018). In this section we summarize the main results obtained from this spectral characterization, describing each family separately.

1. *Polana-Eulalia*. Visible spectra of a total of 65 members of this complex (about 2 % of the family) were obtained using the 10.4m Gran Telescopio Canarias (GTC) and the 3.56m Telescopio Nazionale Galileo (TNG), both located at the El Roque de Los Muchachos Observatory, in the island of La Palma (Spain), and the 3.6m New Technology Telescope (NTT), located at La Silla Observatory (Chile). The results obtained from the spectral analysis of these asteroids are summarized in Fig. 3a and published by de León et al. (2016). Despite the apparent dynamical complexity of this particular region of the asteroid belt, we found no spectral differences between the members of the so-called New Polana and the Eulalia families (Walsh et al. 2013), neither in the visible nor the near infrared (Pinilla-Alonso et al. 2016). An almost equal proportion of B- and C-type asteroids populates this complex, i.e., the family is characterized by asteroids presenting featureless spectra, with slopes ranging in a continuum from blue (-2.9 %/1000Å) to moderately red (2.7 %/1000Å), and with no signs of aqueous alteration. This group is by far the largest primitive collisional family in the inner belt and is also the oldest.

2. *Erigone*. A total of 101 members of this collisional family were observed using the 10.4m Gran Telescopio Canarias, and the results were published by Morate et al. (2016). This set accounts for about 6% of the family, which has 1776 members. Contrary to what is observed in the Polana-Eulalia complex, the Erigone collisional family shows a broader diversity of primitive spectral classes, as is shown in Fig. 3b. The obtained spectral slopes range from blue (-2.8 %/1000Å) to significantly red (7.2 %/1000Å). Perhaps the most remarkable result is the high abundance of asteroids (more than 50%) showing the 0.7 µm absorption band produced by aqueous alteration of silicates, in contrast to the lack of such band found in the Polana-Eulalia complex. The parent body of this family, asteroid (163) Erigone, also presents this hydration feature in its visible spectrum (see Fig. 2). Although the majority (75%) of the asteroids with this absorption band are C-types, as expected, we also find signs of hydration among X-types (15%), B-types (8%), and T-types (2%).

3. *Sulamitis*. We obtained visible spectra of a total of 64 members of the Sulamitis family, accounting for 21% of the total number, using the 10.4m GTC (Morate et al. 2018). As in the case of Erigone, the members of the family showed a significant diversity of primitive spectral classes (see Fig. 3d), distributed in a similar proportion, except B-type asteroids (there is only one object in the sample). Similarly, (752) Sulamitis, like Erigone, presents the 0.7 µm absorption feature in its visible spectrum, and about 60% of the members of the family show this hydration band. Spectral slopes in this family range from moderately blue (-1.6 %/1000Å) to considerably red (8.3 %/1000Å), again similar to what we see in the Erigone family. Also, the two families have comparable ages (see Table 2).

*4. Clarissa*. This group is the smallest family of the ones we present in this work. It has 179 members, and we have obtained visible spectra of a total of 33 (18% of the family) using the 10.4m GTC (Morate et al. 2018). The Clarissa family shares some of the spectral properties of the Polana-Eulalia complex: it is mostly composed of featureless B- and C-type asteroids, although it presents a more significant proportion of redder X-types. Spectral slopes range from blue (-4.4 %/1000Å) to moderately red (4.7 %/1000Å). Interestingly, as in the case of the Polana-Eulalia complex, we find almost no asteroids (3 out of 33) showing the 0.7 µm absorption band (Fig.3c).

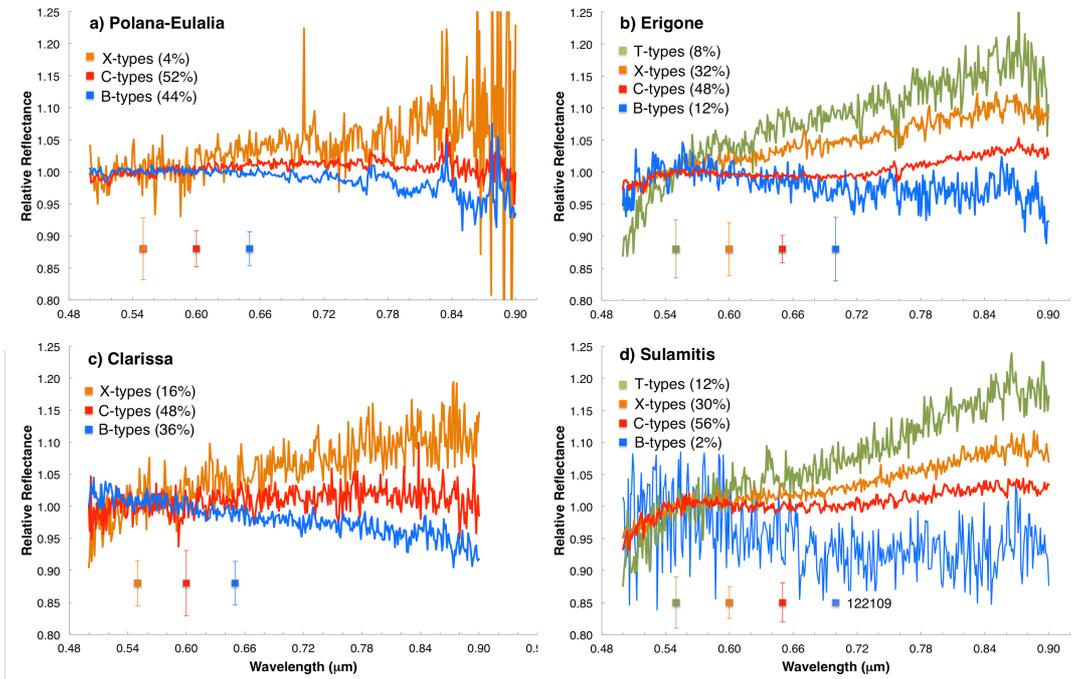

**Figure 3**. Summary of the results obtained from the spectral characterization of the inner belt primitive families studied so far. Each panel contains: the distribution of the taxonomical classes found within the family (percentage); the mean spectrum obtained from the averaging of all the spectra from each class in the family; the corresponding ±1σ deviation from the mean for each class, plotted as error bars at the bottom of each panel. For all the families blue is for B-types, red is for C-types, orange is for X-types and green is for T-types. Spectra are all normalized to unity at 0.55 µm. In the case of the Sulamitis family, there is only one B-type object, asteroid (122109).

Figs. 4 and 5 summarize the spectral characteristics of the primitive families we have analyzed so far. Fig. 4 shows the distributions of the spectral slopes *S'* computed for each family. The vertical lines correspond to the mean slope of each family, shown in Table 3. A simple visual inspection reveals the similarity between the Polana-Eulalia complex and the Clarissa family on one side, and between the Erigone and the Sulamitis families on the other. These similarities were quantified and confirmed by Morate et al. (2018) using a two-sample Kolmogorov-Smirnov test over the two pairs of distributions. In Fig. 5 we have plotted together the distribution (pie charts) of the taxonomical classes of each family.

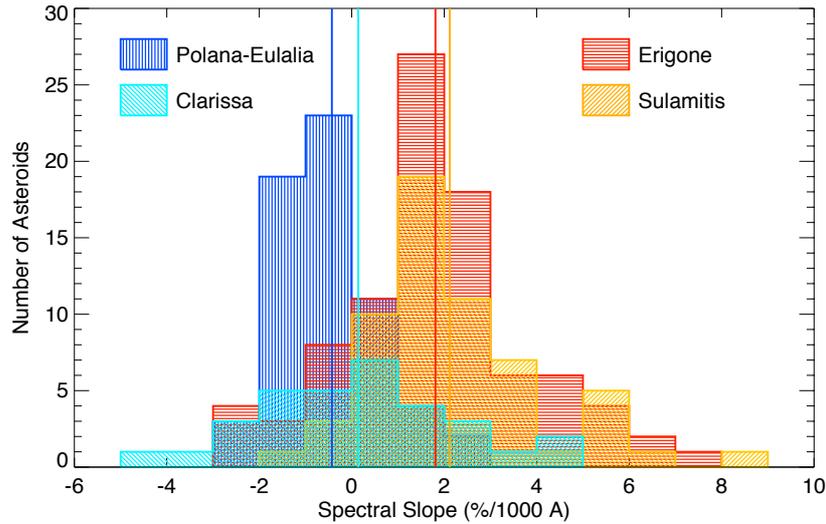

**Figure 4**. Spectral slope distributions of the four inner primitive families studied in this work. Details for each family can be found in de León et al. (2016) for the Polana-Eulalia complex, Morate et al. (2016) for the Erigone family, and Morate et al. (2018) for the Sulamitis and the Clarissa families.

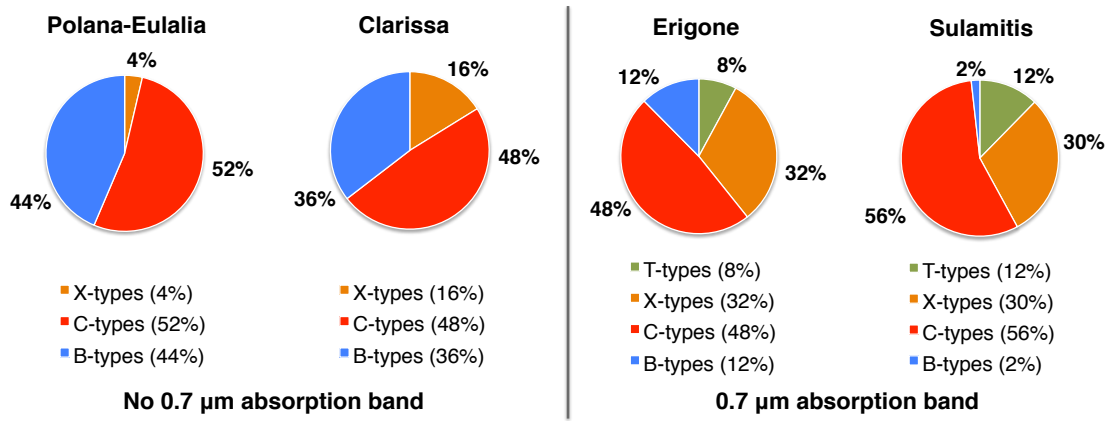

**Figure 5**. Comparison of the distributions of taxonomic classes found for each family. We have grouped the Polana-Eulalia complex and the Clarissa family (left), and the Erigone and Sulamitis families (right) to enhance the similarities between each pair.

We have also computed the average geometric albedo for each family, using WISE/NEOWISE data (Wright et al. 2010; Mainzer et al. 2011), the NEATM model implemented by Alí-Lagoa et al. (2017), and values for $H$ and $G_{12}$ from Oszkiewicz et al. (2012). These average albedos and their associated errors are shown in Table 3, together with the number of asteroids used to compute the average for each family (sample size). We also include in the last column of Table 3 a flag for the quality criteria ($Q$) as defined by Alí-Lagoa et al. (2016), indicating whether the sample includes only those fits having at least 10 data points in both W3 and W4 filters[2] (labeled "Yes" or "No"). The first requirement ensures a reasonable rotational sampling and averaging out of the irregularities in both thermally dominated bands is a requisite to fit the beaming parameter and improve the inferred diameters (Harris 1998). As expected we find average albedo values for the families in good agreement with the albedo values of the parent bodies.

---

[2] The NASA Wide-field Infrared Survey Explorer (WISE) mission is equipped with four filters, W1, W2, W3, and W4 centered at 3.4, 4.6, 12, and 22 μm, respectively.

Table 3. Properties of the four primitive collisional families studied in this work.

| Family | #[1] | Age[2] (Myr) | <S'> (%/1000Å) | <$p_V$>[3] | Sample size | Q[4] |
|---|---|---|---|---|---|---|
| Polana-Eulalia | 3783* | 1400 ± 150 <br> $830^{+370}_{-100}$ | -0.43 ± 1.17 | $0.052^{+0.010}_{-0.009}$ | 615 | Yes |
| Clarissa | 179 | ~60 | 0.14 ± 2.09 | $0.042^{+0.020}_{-0.012}$ | 19 | No |
| Erigone | 1776 | 130 ± 30 | 1.81 ± 2.02 | $0.047^{+0.020}_{-0.013}$ | 212 | Yes |
| Sulamitis | 303 | 200 ± 40 | 2.12 ± 1.83 | $0.053^{+0.010}_{-0.009}$ | 50 | Yes |

[1] From Nesvorny et al. (2015).
[2] From Bottke et al. (2015). The two ages correspond to the so-called "new Polana" (top) and Eulalia (bottom) families, as described in the paper.
[3] Albedo values computed from the NEATM diameters of Alí-Lagoa et al. (2017) to fit WISE/NEOWISE thermal infrared data and the $H$ and $G_{12}$ values from Oszkiewicz et al. (2012).
[4] Q: applied quality criteria based on the number of available data per band defined in Alí-Lagoa et al. (2016).
* K. Walsh, personal communication.

From the overall spectral characteristics shown in Figs. 3-5, we can roughly differentiate two groups in the inner belt primitive families studied so far: the Polana-like group and the Erigone-like group. The Polana-like family members present featureless (no 0.7 μm absorption band), homogeneous spectra ranging from slightly blue to moderately red. The Erigone-like group contains members showing a more extensive spectral diversity and, in their majority, the 0.7 μm band associated with phyllosilicates. While Erigone and Sulamitis families have similar ages and mean albedo values, the Clarissa family is much younger than the Polana-Eulalia complex (see Table 3). Also, we find a more substantial fraction of red, X-type asteroids in the Clarissa family (Fig. 5). Recent laboratory experiments to simulate space weathering effects on low-albedo, primitive materials suggest that their visible spectra tend to get bluer and their albedo tend to get higher as exposure age increases (Lantz et al. 2015, 2017). This result can explain the fact that the younger Clarissa family presents a lower fraction of B-types (blue) and a larger fraction of X-types (red) than the much older Polana-Eulalia complex (Campins et al. 2018). Regarding the albedo, we cannot establish any conclusion, as the number of asteroids with good-quality albedo values in the Clarissa family is too low.

## 3. Comparison with the spectra of Bennu and Ryugu

Primitive near-Earth asteroids (101955) Bennu and (163172) Ryugu are the targets of NASA's OSIRIS-REx and JAXA's Hayabusa2 sample-return missions, respectively. These two missions are currently in space and are expected to arrive at their targets during the second half of 2018. In both cases, our results are especially relevant to the planning and interpretation of in situ images and spectra to be obtained by the two spacecraft. Spectra and color images will not only provide information on the visible spectral characteristics of the surface of the asteroid but will be used to identify and select the site for the sample collection. Therefore, it is essential to have as much information as possible on what to expect regarding colors in the visible wavelength region of the surface of the two asteroids before the arrival. This is only possible from the examination of ground-based spectra of the two targets, and also from the visible spectra of the asteroids located in the inner belt, their most likely source region.

### 3.1. (101955) Bennu: target of the NASA OSIRIS-REx mission

Primitive NEA (101955) Bennu (previously known by its provisional designation 1999 RQ36) is the target of the NASA's OSIRIS-REx sample-return mission (Lauretta et al. 2015; Lauretta et al. 2017). This asteroid has a diameter of about 500 m, a surface with an overall blue color (B-type asteroid) and a low albedo ($p_V$ = 0.04). It has been observed only two times in the visible wavelength region using ground-based telescopes. The spectrum shown in Fig. 6 corresponds to the average of the visible spectra obtained over five consecutive nights between 15 and 20 September 1999, during one of Bennu's close approaches to the Earth. Observations span a range of phase angles from 35.9° to 65.0° and visual magnitudes from 16.6 to 14.4. Visible colors (black squares in Fig. 6) were obtained using *ubvwxp* ECAS filters on September 14-17 2005 (Hergenrother et al. 2013). Both B-type and F-type asteroids show a turnover in reflectance between 0.4 and 0.5 µm, with the B-types presenting a larger decrease in reflectance short ward 0.5 µm. This is due to absorption in the ultra-violet (UV) wavelength region associated with the presence of aqueously altered minerals. The ECAS colors of Bennu do not show this UV drop-off in reflectance, and the only published visible spectrum of Bennu starts at 0.47 µm, so we cannot detect this drop-off. Interestingly, neither the spectrum nor the colors show the 0.7 µm absorption band associated with hydrated silicates (phyllosilicates). This was first noted by Hergenrother et al. (2013) and later studied by de León et al. (2016). There are several published spectra of Bennu in the near-infrared wavelength region (0.8-2.4 µm) that range from negative, blue spectral slope (-0.30 %/1000Å) to positive, red slope (1.40 %/1000Å), i.e., from a B-type to a C-type according to the DeMeo et al. (2009) taxonomical classification. This variation is compatible with having a negative, B-type slope in the visible, as shown by de León et al. (2012), and it was thoroughly studied by Binzel et al. (2015). The authors found no correlation between near-infrared slope and any systematic observational effect (including phase angle), and so, they propose an alternative explanation related to the accumulation of finer grained material in an equatorial ridge created by regolith migration during episodes of rapid rotation.

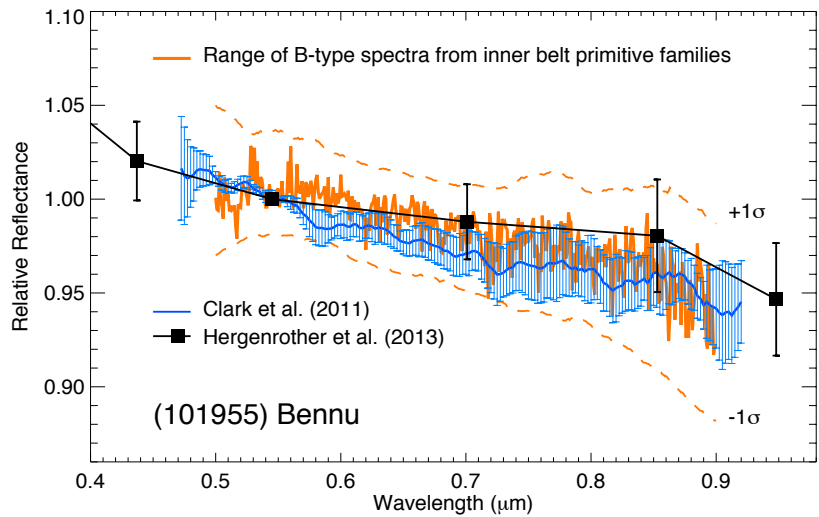

**Figure 6**. Visible spectrum and colors for primitive NEA Bennu, target of the NASA OSIRIS-REx mission. We show here the mean spectrum of all the B-type asteroids found among the inner belt primitive families, as well as the corresponding ±1σ deviation of the mean.

We showed in Fig. 8 from de León et al. (2016) that the visible spectrum of Bennu ($S'$ = -1.28 ± 0.03 %/1000Å) is almost identical to the visible spectrum of (142) Polana. Bennu also lies within the boundaries defined by the ±1σ of the mean spectrum of the Polana-Eulalia family members ($S'$= -0.43 ± 1.17 %/1000Å). Here we have computed the mean spectrum of all the asteroids classified as B-types in the four primitive families studied in this work. The resulting mean spectrum and the corresponding ±1σ deviation from the mean are shown in Fig. 6 (orange), together with the spectrum and the colors of Bennu. As expected, the agreement is excellent.

Regarding the origin of Bennu, spectroscopic and dynamical arguments suggest that it most likely originated in the primitive collisional families of the inner belt, in particular in the Polana-Eulalia complex (Campins et al. 2010; Bottke et al. 2015). The absence of a 0.7 µm absorption band in the existing spectra of Bennu is additional evidence in favor of an origin in the Polana-Eulalia complex (Campins et al. 2018). Nevertheless one should also consider the possibility that Bennu has lost the 0.7 µm band due to its proximity to the Sun as a near-Earth asteroid or that space weathering might have as well removed the signs of such band. Regarding this last point, a work by Matsuoka et al. (2015) shows that space weathering effects on C-type asteroids include a diminishing in the depth of the 0.7 µm band; on the other hand, Lantz et al. (2018) state that the effects of space weathering on the 0.7 µm absorption band have not been deciphered yet through laboratory experiments.

### 3.2. (163172) Ryugu: target of the JAXA Hayabusa2 mission

Primitive NEA (162173) Ryugu (previously known by its provisional designation 1999 JU3) is the main target of the Japanese sample-return mission Hayabusa2 (Tsuda et al. 2013). It is a small (about 800 m), low-albedo ($p_V$ = 0.05 – 0.07; Campins et al. 2009, Müller et al. 2017) asteroid classified as a C-type and thoroughly observed from ground-based telescopes, in particular in the visible wavelength range. Fig. 7 shows all the available visible spectra of Ryugu. When more than one spectrum have been obtained, we show the mean spectrum and its corresponding standard deviation (error bars). This is the case for spectra labeled as L12 (Lazzaro et al. 2013), S12 (Sugita et al. 2013), M12 (Moskovitz et al. 2013), P16F (Perna et al. 2017, using FORS2), and P16X (Perna et al. 2017, using XShooter). Table 4 summarizes the observational circumstances for all the spectra, including the observation date and the apparent visual magnitude ($m_V$) and phase angle (α) at the time of observation. We have also included in Table 4 the taxonomical classification of each spectrum, as well as our computed spectral slope $S'$ in the 0.55-0.90 µm wavelength range. We do not find any clear correlation between the spectral slope and the phase angle nor the aspect angle, and so the apparent differences between the available spectra might be explained by the use of different instruments/telescopes, observing and calibration issues, use of various solar analogs, etc. Nevertheless, the variation falls within the error bars of the data. Several attempts have been made to search for surface heterogeneity in Ryugu (Lazzaro et al. 2013; Moskovitz et al. 2013; Perna et al. 2017), with no conclusive results. In the near-infrared wavelength range, published spectra by Abe et al. (2008), Pinilla-Alonso et al. (2013) and Perna et al. (2017) are similar, showing a neutral to slightly red spectral slope, while the most recent near-infrared spectra by LeCorre et al. (2018) presents a significantly redder slope. These authors fail to find a plausible explanation for such discrepancy.

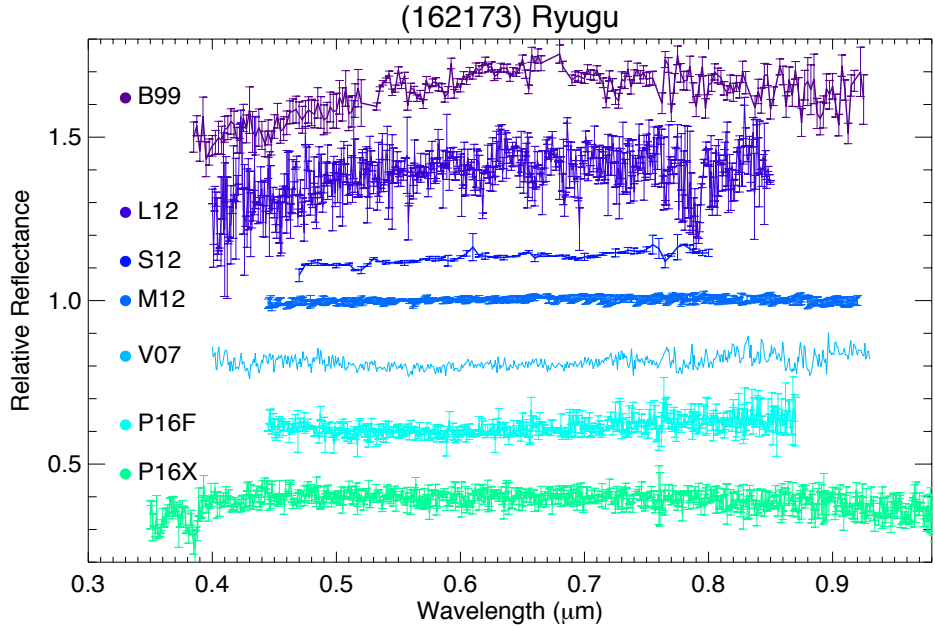

**Figure 7**. Visible spectra of primitive NEA (162173) Ryugu, target of the JAXA Hayabusa2 mission. This asteroid has been thoroughly observed from the ground (see main text for data sources).

Some of the visible spectra of Ryugu led to a taxonomical classification as Cg. Asteroids belonging to this spectral class show a drop in reflectance short ward of 0.55 µm, due to absorption in the ultra-violet wavelength region associated with the presence of aqueously altered minerals. Only one visible spectrum of Ryugu, obtained by Vilas (2008) presented an absorption band at 0.7 µm produced by hydrated silicates (this spectrum had poor signal-to-noise). The rest of the available spectra, shown in Fig. 7, do not illustrate this absorption. However, it is important to emphasize that the absence of the 0.7 µm band does not imply the lack of hydrated minerals. In roughly half of the studied cases, asteroids showing an absorption band in the 3-µm region due to hydrated silicates do not show the corresponding band at 0.7 µm (Vilas 1994; Rivkin et al. 2002, 2015). A recent study made by Busarev et al. (2018) based on the shape of visible spectra of Ryugu from Vilas (2008) and Sugita et al. (2013) obtained a month after aphelion passage, suggests the existence of sublimation/degassing activity of Ryugu and the presence of an ice reservoir at small depths, indicative of a relatively short residence time in the near-Earth space.

**Table 4**. Summary of the observational circumstances of the available visible spectra of Ryugu.

| Data ID | Obs. Date | $m_V$ | α (°) | Tax. | S' (%/1000 Å) |
|---|---|---|---|---|---|
| B99 | 17/05/1999 | 17.7 | 6.1 | Cg | -0.75 ± 0.20 |
| V07 | 10/09/2007 | 17.9 | 22.5 | Cb | 0.93 ± 0.07 |
| M12 | 1-3/06/2012 | 17.7-17.9 | 0.2-2.0 | C | -0.02 ± 0.02 |
| S12 | 24-26/06/2012 | 19.1-19.6 | 22.7-30.3 | C | 1.38 ± 0.11 |
| L12 | 9-10/07/2012 | 19.9 | 33.0 | Cg | -0.06 ± 0.23 |
| P16F | 12/07/2016 | 18.9 | 13.9 | Cb | 1.52 ± 0.09 |
| P16X | 11/08/2016 | 19.3 | 24.5 | C | -0.37 ± 0.06 |

B99 – Binzel et al. (2001); V07 – Vilas (2008); M12 – Moskovitz et al. (2013); S12 – Sugita et al. (2013); L12 – Lazzaro et al. (2013); P16F – Perna et al. (2017), using FORS2; P16X – Perna et al. (2017), using XShooter

As in the case of Bennu, different studies indicate that Ryugu's most likely origin is in the primitive collisional families of the inner belt or the low-albedo and low-inclination population of background asteroids in the same region (Campins et al. 2013; Bottke et al. 2015). To further support this idea, we have taken the most recently obtained visible spectra of Ryugu from Perna et al. (2017) and compared them with the mean spectra of all the B-types and all the C-types found among the four primitive families studied in this work (Polana-Eulalia, Erigone, Sulamitis, and Clarissa). The two selected spectra of Ryugu represent very well the range of variation in spectral slope that is observed among all the available spectra (Table 4). Fig. 8 shows a comparison between these two spectra of Ryugu and the mean spectra obtained for all the B-types (orange) and C-types (red) find among the four primitive families presented in this work. The spectrum labeled as P16X corresponds to a visible-to-near-infrared (VNIR) spectrum obtained using X-Shooter (Perna et al. 2017), i.e., although the visible part presents a negative spectral slope (B-type), the complete VNIR spectrum corresponds to a C-type according to the DeMeo et al. (2009) taxonomy. As in the case of Bennu, the absence of a 0.7 μm absorption band in the spectra of Ryugu obtained to date is additional evidence in favor of an origin in the Polana-Eulalia complex (Campins et al. 2018).

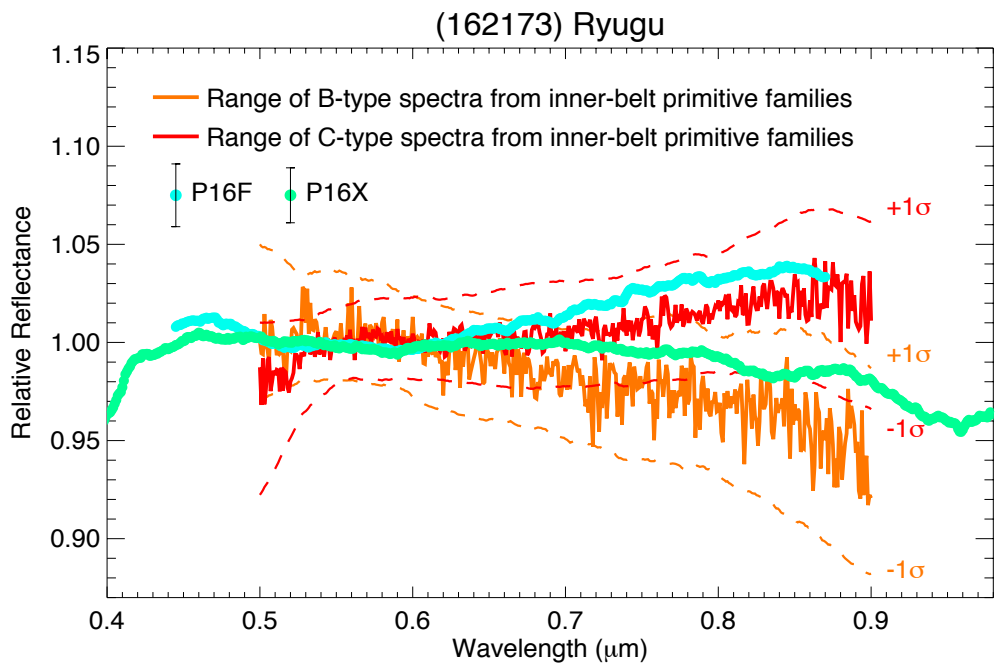

**Figure 8**. Visible spectra of primitive NEA Ryugu, target of the JAXA Hayabusa2 mission. We show here the mean spectra of all the B-type (orange) and C-type (red) asteroids found among the inner belt primitive families, as well as the corresponding ±1σ deviation of the mean. Visible spectra of Ryugu are from Perna et al. (2017) and have been smoothed to a factor of 50 for a better visualization. P16F corresponds to the spectra obtained using FORS2 instrument, while P16X corresponds to those obtained using XShooter.

## 4. Discussion

*(101955) Bennu*. The only published visible spectrum of asteroid Bennu presents a blue, negative spectral slope, similar to the visible spectrum of B-type asteroid (142) Polana. This spectrum is compatible with the mean of the members of the Polana-Eulalia complex and, in general, consistent with the mean spectrum of all the B-type asteroids found within the four collisional families of primitive asteroids studied so far: Polana-Eulalia, Erigone, Sulamitis, and Clarissa. Dynamical simulations indicate that the most likely origin of Bennu is the Polana-Eulalia complex (Bottke et al. 2015). Therefore, assuming this origin, and considering the visible spectral slopes computed for this family in de León et al. (2016), we can expect a variation in the slope of $S'$ = [-1.64, 0.70] %/1000Å. When considering only members of the Polana-Eulalia complex classified as B-types, the expected variation would be $S'$ = [-1.83, -0.84] %/1000Å. Being even more conservative and including also B-type asteroids from the other studied inner-belt primitive families as potential sources of Bennu, we can expect a variation of $S'$ = [-2,28, -0.78] %/1000Å. These ranges have been computed as [$<S'>+1\sigma$, $<S'>-1\sigma$] using the mean values $<S'>$ and their corresponding $1\sigma$ errors shown in Table 5. Note that we have computed the average of the spectral slopes and not the spectral slope of the average spectra.

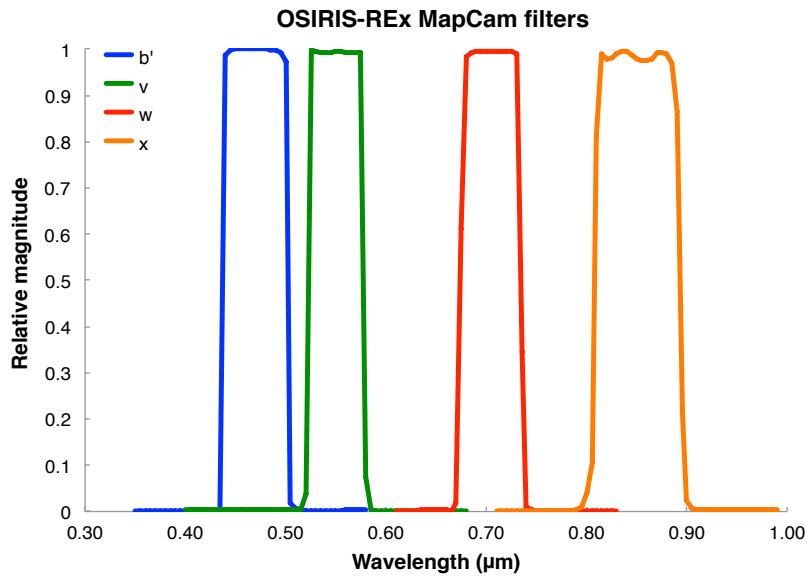

**Figure 9**. OCAMS MapCam filter transmission curves. Figure modified from Rizk et al. (2018).

In August 2018 the OSIRIS-REx mission will begin the approach phase to asteroid Bennu and start obtaining images as a point source using its suite of optical cameras. Resolved images will be obtained starting in late September 2018. The OSIRIS-REx Camera Suite (OCAMS) includes three cameras: SamCam, MapCam, and PolyCam. MapCam is equipped with a filter wheel and a set of 6 filters: 2 clear (panchromatic) filters and four color filters, aligned with the ECAS wavelengths (*b'vwx*) covering a wavelength range between 0.4 and 0.9 µm, approximately (see Fig. 9). Filter *b'* has been shifted toward longer wavelengths from the ECAS blue filter to improve optical and radiometric performance and minimize aging effects due to radiation (Rizk et al. 2018). Color images are fundamental for assessing surface composition of the target, and to subsequently select the best sample-collection site. Also, MapCam colors will be used to assess the extent of space weathering on Bennu and to help in the determination of surface "freshness", which will have higher scientific value for the sampling decision. The calculation of the *b'/v* and *v/x* color ratio maps is stipulated in two of the mission requirements documents in order to characterize the UV slope, and the visible slope, respectively, as well as the presence and depth of any 0.7 µm feature.

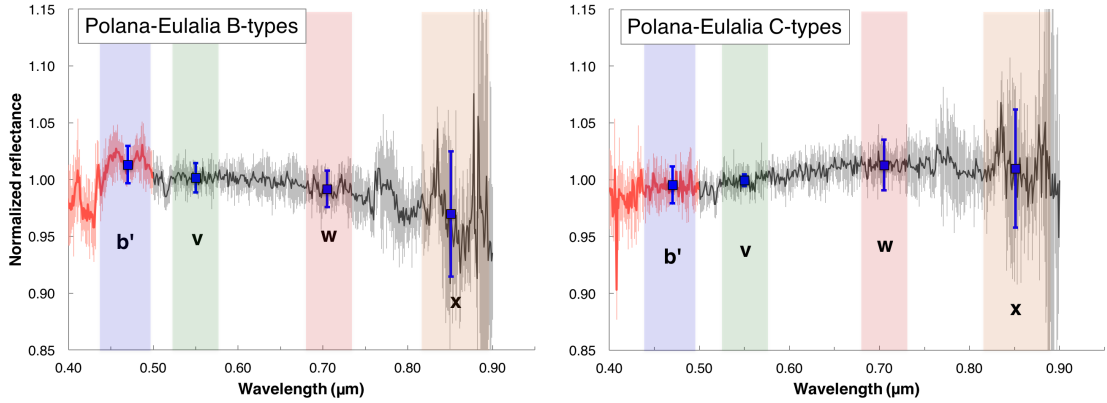

**Figure 10**. Example of the obtained reflectances and their corresponding errors for the OCAMS MapCamp filters (*b'vwx*) after the convolution of their transmission curves and the represented spectra. These are the mean spectra of all the B- (left panel) and C-type (right panel) asteroids in the Polana-Eulalia complex (black) and the shadowed grey region corresponds to the standard deviation of the mean (de León et al. 2016). The red part of these spectra correspond to the mean spectrum of those B-type (left) and C-type (right) asteroids in the Polana-Eulalia complex observed down to 0.35 μm, and used to compute the *b'* reflectance (see main text for details). Reflectances and mean spectra have been normalized to unity at the central wavelength of *v* filter for comparison. Colored regions correspond to the width of the different filters.

Using the OCAMS MapCam transmission curves of the *b'vwx* filters, we have computed the range of (*b'-v*) and (*v-x*) colors expected from the visible spectra of the different populations considered as representative of the surface composition of Bennu and listed in Table 5. These include the B-type and C-type asteroids in the Polana-Eulalia complex, and B-type and C-type asteroids from the four collisional families of primitive asteroids presented in this work: Polana-Eulalia, Erigone, Sulamitis, and Clarissa. To obtain these colors we have computed the integral of the average spectra of these populations through the OCAMS MapCam filters using the relation

$$r_i = \frac{\int_{\lambda=\lambda_1}^{\lambda_2} f_i(\lambda) r(\lambda) \lambda d\lambda}{\int_{\lambda=\lambda_1}^{\lambda_2} f_i(\lambda) \lambda d\lambda},$$

where $r_i$ is the reflectance in the *i*-th filter, λ is the wavelength, $f_i(\lambda)$ is the filter transmission function, $r(\lambda)$ represents the asteroid reflectance spectra, and $\lambda_1$ and $\lambda_2$ are the lower and upper bounds of the filter transmittance, respectively. Note that, as we are working with reflectance spectra (relative to the Sun) the instrumental terms in the above relation, like the detector QE or the camera optics reflectivity, are cancelled. We apply this relation to the mean spectrum of each population, considering the ±1σ deviation associated with this mean spectrum to compute the errors in reflectances (see Fig. 10). The filter reflectance and its associated error are given by the average and the variance of these values. Finally, we obtain the colors using the relation

$$m_i - m_j = -2.5 \times \log_{10}(r_i/r_j),$$

where $r_i$ and $r_j$ are the reflectances at the *i*-th and *j*-th filters and the color *i-j* is the difference in the magnitudes $m_i$ and $m_j$. These colors are directly comparable to ECAS colors since they use a solar spectrum as the reference for the magnitude system. Fig.

10 shows an example of the obtained reflectances through the OCAMS filters for the mean spectrum of all the B-type asteroids in the Polana-Eulalia complex (black line). Note that in this case and also for the rest of the populations, we do not have data below 0.5 μm. To compute the reflectance through the *b'* filter (and so the *b'-v* color) in this particular case, we used spectra of a sample of B-type asteroids in the Polana-Eulalia complex obtained using the 3.6m Telescopio Nazionale Galileo (TNG) and New Technology Telescope (NTT). These spectra go down to 0.35 μm (see de León et al. 2016). We computed the mean spectrum of a total of 16 B-types and used this mean spectrum to obtain the *b'* reflectance (red line in Fig. 10). The same approach has been followed with a sample of 9 C-type asteroids in the Polana-Eulalia complex. For the rest of the populations listed in Table 5 we used visible spectra of B- and C-type asteroids from the SMASS database (0.43 – 0.92 μm), all of them located in the same region of the belt as that of the 4 primitive families studied here (Bus & Binzel 2002). The obtained *b'-v* and *v-x* colors are shown in Table 5 and the regions covered by the different populations in the color-color plot considering the errors in the computed colors are shown in Fig. 11. We have also plotted the *b'-v* and *v-x* colors computed for Bennu by Hergenrother et al. (2013).

**Table 5**. Summary of the average slopes and their corresponding 1σ errors for the different populations presented in this work. We also include the *b'-v* and *v-x* colors computed using the OCAMS MapCam filters transmission curves (see main text for details). The visible spectral slope of Bennu has been taken from de León et al. (2016) and colors from Hergenrother et al. (2013).

| Population | $\langle S' \rangle$ (%/1000Å) | *b'-v* | *v-x* |
|---|---|---|---|
| Polana-Eulalia B-types | -1.34 ± 0.49 | -0.0124 ± 0.0225 | -0.0348 ± 0.0626 |
| Polana-Eulalia C-types | -0.10 ± 0.87 | 0.0043 ± 0.0184 | 0.0113 ± 0.0573 |
| Primitive families B-types | -1.53 ± 0.75 | 0.0053 ± 0.0256 | -0.0540 ± 0.0194 |
| Primitive families C-types | 0.61 ± 0.76 | 0.0379 ± 0.0300 | 0.0247 ± 0.0132 |
| Bennu | -1.28 ± 0.03 | -0.03 ± 0.02 | -0.01 ± 0.03 |

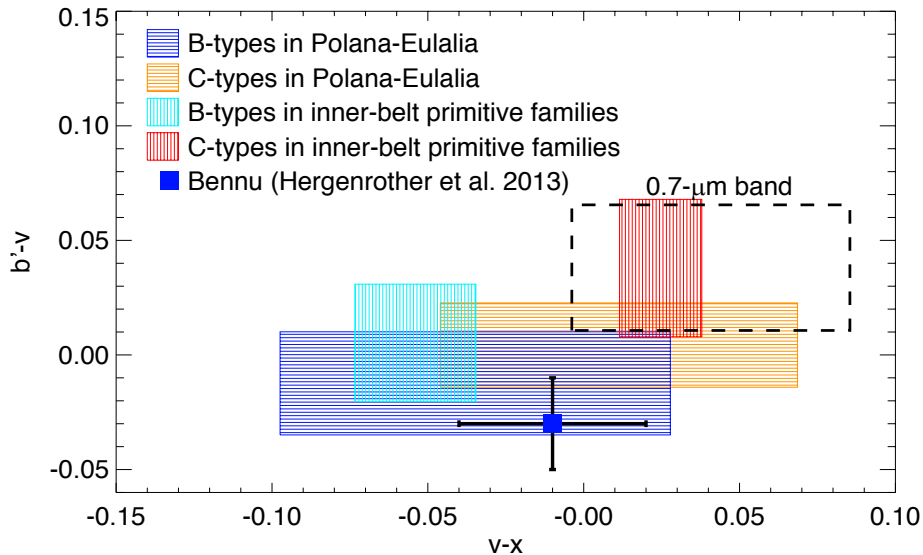

**Figure 11**. Expected color variation computed for the MapCam filters on OSIRIS-REx and the mean spectra of B- and C-type asteroids from the Polana-Eulalia complex only and from the four collisional families of primitive asteroids in the inner belt: Polana-Eulalia, Erigone, Sulamitis, and Clarissa. See main text for details on the color computation. The dashed-line region corresponds to the color variation computed from the asteroids in the primitive families showing the 0.7 μm absorption band.

As we can see in Fig. 11, the color regions associated with the B- and C-type asteroids in the Polana-Eulalia complex present a larger dispersion in the (*v-x*) color than in the (*b'-v*) color. This is expected, as the (*v-x*) represents the visible slope and what we found in the Polana-Eulalia complex was a continuum of spectral slopes, from blue to moderately red (de León et al. 2016). Also, the data obtained by these authors presented higher levels of noise as the spectra approached 0.9 µm (*x* filter, see Fig. 10). If one considers the colors obtained from the mean B- and C-type spectra of the four collisional families of primitive asteroids in the inner belt, we can easily separate between the two taxonomies. The colors of Bennu fall in the region outlined by the B-type asteroids in the Polana-Eulalia complex, as expected from their visible spectra.

Although neither the spectrum of Bennu nor the spectra of any of the Polana-Eulalia complex present the 0.7 µm absorption band, we do not rule out the possibility of finding such band when observing the resolved surface of the asteroid. Therefore, we have computed the (*b'-v*) and (*v-x*) colors on the mean spectra of all the asteroids in the inner belt families showing the 0.7 µm absorption band, mainly members of the Erigone and Sulamitis families. As we did for the other populations, we computed the mean spectra of Ch-type asteroids from the SMASS database and located in the inner part of the main belt, to obtain the *b'* reflectance value and so the *b'-v* color. The region covered by the two colors and their corresponding errors is depicted with a dashed-line in Fig. 11. We have also computed the band depth on both the mean spectrum and the spectrum of the asteroid showing the deepest absorption band. Band depth has been calculated following the procedure described in De Prá et al. (2018). We fit a straight line (continuum) to the reflectances at the *v* and *x* filters; divide *v*, *w*, and *x* reflectances by this line (continuum removal); compute the band depth as $1 - R_w$, where $R_w$ is the new reflectance at the *w* filter after continuum removal. According to the obtained results, if the 0.7 µm band is detected, we can expect it to have depths of the order of 2.1 ± 3.5 % on average, with the deepest absorption being 7.6 ± 2.0 %.

*(162173) Ryugu*. Hayabusa2 will arrive at Ryugu about one month before OSIRIS-REx encounters Bennu. The spacecraft is provided with the Optical Navigation Cameras (ONC). The ONC consists of a narrow-angle camera (ONC-T, Kameda et al. 2017), equipped with a filter wheel and a set of 7 filters, covering the wavelength range from 0.4 µm to 0.95 µm, and two wide-angle cameras (ONC-W1 and ONC-W2).

In the case of asteroid Ryugu, there are several published visible spectra, as detailed in the previous section. Although the majority of them are in agreement with a taxonomical classification of a C-type, there is one recently obtained spectrum that presents a negative spectral slope in the visible and shows the concave-up shape typical of C-types in the near-infrared according to DeMeo et al. (2009). Again, the most likely origin of Ryugu from dynamical simulations is the Polana-Eulalia complex, although the low-albedo, low-inclination background population has also been considered (Campins et al. 2010). Assuming an origin in the Polana-Eulalia complex one can expect the same variation in spectral slope as that for Bennu and indicated above. Considering only those members of the family complex classified as C-type asteroids, then *S'* = [-0.97, 0.77] %/1000Å. If one also considers the possibility of Ryugu coming from any of the four primitive families in the inner belt studied so far, and only from the C-type asteroids, we can expect a variation in slope of *S'* = [-0.15, 1.37] %/1000Å. This is in good agreement with the observed variation in the spectral slope of Ryugu, *S'* = [-0.75,1.52] %/1000Å.

For both Ryugu and Bennu, the possibility remains of finding more spectral variation in the visible spectra as surface resolved measurements are acquired. For the two targets of the space missions, the best spectral variation range to consider is the one computed from all B-types and C-types studied so far in the primitive collisional families of the inner belt. Table 5 summarizes these results.

## 5. Conclusions

The latest results obtained within the frame of our PRIMitive Asteroids Spectroscopic Survey (PRIMASS) have allowed us to characterize the spectral behavior in the visible wavelengths of 4 collisional families of primitive asteroids located in the inner belt: the Polana-Eulalia complex and the Erigone, Sulamitis, and Clarissa families. This region, and in particular the Polana-Eulalia complex, is considered the most likely origin of the two targets of the current sample-return missions: NASA's OSIRIS-REx (Bennu) and JAXA's Hayabusa2 (Ryugu). In this paper, we have performed a global analysis on the spectral characteristics of the members of these families and have compared them to the available visible spectra of Bennu and Ryugu. The results can be summarized as follows:

- According to the visible spectra obtained from their members, we can differentiate two groups among the families of primitive asteroids studied so far. The Polana-like group presents homogeneous, featureless spectra in a continuum of slopes from blue to moderately red, and no 0.7 μm band. The Erigone-like group, shows a spectral diversity among primitive taxonomic classes and a majority of spectra with the 0.7 μm band associated with phyllosilicates.
- The obtained spectral slopes of the inner-belt families of primitive asteroids indicate that we should expect variations in slope between -2.28 and -0.78 %/1000 Å for Bennu and between -0.15 and 1.37 %/1000 Å for Ryugu.
- While the Erigone and the Sulamitis families show very similar properties regarding spectra, collisional ages, and visible albedo, we see that the Clarissa family has on average a redder spectral slope and is significantly younger than the Polana-Eulalia complex. A tentative explanation for these differences is space weathering. Laboratory experiments with carbonaceous chondritic material show that lowest-albedo material gets bluer with space exposure age (Lantz et al. 2015, 2017). This has testable implications for Bennu and Ryugu, where older terrains would be expected to be bluer than younger surfaces (Campins et al. 2018).
- The mean spectra of B- and C-type asteroids in the studied inner-belt families through the OCAMS filters show that we will be able to differentiate between B- and C-type surfaces (blue and red) using the *b'-v* and *v-x* colors. In the event of detecting the presence of the 0.7 μm absorption band, we expect to find band depths of 2.1 ± 3.5 % on average, with maximum absorption of 7.6 ± 2.0 %.

## Acknowledgements

We want to especially thank Dr. Davide Perna for sharing his spectral data from Ryugu and Dr. Sonia Fornasier for her constructive and extremely useful review. JdL acknowledges financial support from the Spanish Ministry of Economy and Competitiveness (MINECO) under the 2015 Severo Ochoa Program MINECO SEV-2015-0548. HC acknowledges support from NASA's Near-Earth Object Observations program and the Center for Lunar and Asteroid Surface Science funded by NASA's SSERVI program at the University of Central Florida. DM gratefully acknowledges MINECO for the financial support received in the form of a Severo Ochoa Ph.D. Fellowship. JdL, JL, MP, and JLR acknowledge support from the project AYA2015-67772-R (MINECO). The research leading to these results has received funding from the European Union's Horizon 2020 Research and Innovation Programme, under Grant Agreement no 687378. This material is based partly upon work supported by NASA under Contract NNM10AA11C issued through the New Frontiers Program.

Pre-print submitted to Icarus - Accepted